# Stark Effect of Hybrid Charge Transfer States at Planar ZnO/Organic Interfaces

Ulrich Hörmann[1*], Stefan Zeiske[1], Fortunato Piersimoni[1], Lukas Hoffmann[2], Raphael Schlesinger[3], Norbert Koch[3], Thomas Riedl[2], Denis Andrienko[4], and Dieter Neher[1*]

[1]Institute of Physics and Astronomy, University of Potsdam, Karl-Liebknecht-Straße 24-25, 14476 Potsdam, Germany

[2]Insitute of Electronic Devices, University of Wuppertal, Rainer-Gruenter-Str. 21, 42119 Wuppertal, Germany

[3]Institut für Physik & IRIS Adlershof, Humboldt-University of Berlin, Brook-Taylor-Straße 6, 12489 Berlin, Germany

[4]Max Planck Institute for Polymer Research, Ackermannweg 10, 55128 Mainz, Germany

* ulrich.hoermann@uni-potsdam.de,
  neher@uni-potsdam.de

## Abstract

We investigate the bias-dependence of the hybrid charge transfer state emission at planar heterojunctions between the metal oxide acceptor ZnO and three donor molecules. The electroluminescence peak energy linearly increases with the applied bias, saturating at high fields. Variation of the organic layer thickness and deliberate change of the ZnO conductivity through controlled photo-doping allow us to confirm that this bias-induced spectral shifts relate to the internal electric field in the organic layer rather than the filling of states at the hybrid interface. We show that existing continuum models overestimate the hole delocalization and propose a simple electrostatic model in which the linear and quadratic Stark effects are explained by the electrostatic interaction of a strongly polarizable molecular cation with its mirror image.

## I. Introduction

Due to the comparably low dielectric constant of organic semiconductors, photo-excited states are strongly bound and spatially localized in these materials. To split these excitons into free charges, organic photovoltaic devices combine donor and acceptor materials with a suitable energy level alignment.[1–4] In such a solar cell, a charge transfer (CT) state is formed at the donor/acceptor interface.[5–8] As generation and recombination of charges proceed though this state, its energy determines the open-circuit voltage $V_{oc}$ and hence the efficiency of the device.[5,9–13]

In hybrid inorganic/organic heterojunctions either the donor or the acceptor is replaced by an inorganic semiconductor.[14–19] By using sensitive photocurrent and electroluminescence (EL) measurements in combination with photoelectron spectroscopy, Piersimoni et al. proved that a hybrid charge transfer (HCT) state is involved in charge generation and radiative recombination at the hybrid junction and that its energy correlates with the interfacial energy level alignment.[20] Shortly after, studies on related systems revealed a distinct blue-shift of the HCT EL spectrum with the applied bias – an effect which is typically not reported in the field of organic photovoltaics.

Eyer et al. proposed that HCT states are confined in a one-dimensional triangular quantum well for their studied ZnMgO/poly (3-hexylthiophene) (P3HT) system.[21,22] The increasing steepness of the linear potential in the photoactive material with larger applied electric field confines the wave function of the charges and hence lifts the ground state energy of the HCT state. Despite its simplicity, this approach reproduced the 2/3- power law dependence of the EL maximum with bias they observed.

Panda et al. investigated the molecular semiconductor 4,4'-bis(N-carbazolyl)-1,1'-biphenyl (CBP) in a heterojunction with sputtered ZnO.[23] They argued that the EL signal originates from a trapped HCT state where the electron is highly localized in a trap state at the ZnO surface. The energy of the HCT emission is then determined by the location of the Fermi-level within the density of trap states. The authors propose that an increasing voltage drop accross the ZnO with larger applied bias shifts the ZnO Fermi-level. This leads to a filling of trap states in the ZnO, accompanied by a blue shift of the HCT.

In the present work we study the donor/acceptor interface in hybrid ZnO/small molecule heterojunction devices with particular focus on the bias dependence of the HCT emission. Our results confirm a blue shift of the HCT peak with increasing bias. Investigation of samples with different organic layer thickness and deliberate modulation of the ZnO conductivity through controlled photo-doping allow us to rule out state filling as the reason for the bias-dependent emission properties, provided that the devices have undergone an initial light soaking procedure. Instead, we find that the electric field across the organic layer determines the HCT state energy at the ZnO/organic interface independent of the interfacial energy alignment. Different quantum mechanical and classical electrostatic models are tested to reproduce the effect. Best agreement with the experimental data is achieved when considering the Stark effect of a highly polarizable molecular cation placed above a heavily n-doped ZnO surface.

## II. Experimental methods

In this study we combine ZnO with three organic molecules with different ionization energies (IE): BF-DPB (N4,N4'-Bis(9,9-dimethyl-9H-fluoren-2-yl)-N4,N4'-diphenylbiphenyl-4,4'-diamine) with an IE of 5.25 eV[24], $\alpha-$NPD ((N,N'-Di(1-naphthyl)-N,N'-diphenyl-(1,1'-biphenyl)-4,4'-diamine) with about 5.4 eV[25] and BPAPF (9,9-Bis[4-(N,N-bis-biphenyl-4-yl-amino)phenyl]-9H-fluorene) with an IE of ca. 5.6 eV[26]. The cartoon in Figure 1 shows the ZnO conduction band minimum energy of around 4.1 eV [27] and the ionization energies of BF-DBP, $\alpha-$NPD and BPAPF for comparison. These organic semiconductors are typically employed as hole transporting materials.[28–30] Here we used them as donors because they form appropriate HCT energy gaps with ZnO in a spectral region that is both detectable with established equipment and well separated from the organic bulk emission.

The configuration of our hybrid devices is a planar stack of ITO (140 nm)/ZnO/organic/MoO$_3$ (13 nm)/Al (100 nm). The devices were fabricated on glass substrates coated with patterned indium tin oxide (ITO) (Luminescence Technology Corp.). The substrates were sonicated in acetone, detergent, milliQ water, and isopropanol and dried under nitrogen flow. Zinc oxide was deposited from solution (SolGel) or by atomic layer deposition (ALD), to investigate whether the preparation route would influence the HCT emission features. The SolGel-ZnO layer was prepared by spincoating a Zinc acetate dihydrate (Sigma Aldrich) precursor solution on the substrate and annealing at 200 °C in air for 1 h.[31] The thickness of the SolGel-ZnO layer is 30 nm. ALD-ZnO was deposited in a commercial Beneq TFS 200 system (base pressure 1.5 mbar). As precursors, diethylzinc and water, both kept at room temperature, were used.[32,33] At a substrate temperature of 80 °C the growth rate per cycle was 0.15 nm. 25 nm of zinc oxide (170 cycles) were deposited on ITO substrates. The substrates were masked by Kapton tape to avoid shunting between the contact of the top electrode and the zinc oxide layer. The ALD-ZnO films were prepared in Wuppertal and express shipped in air to Potsdam for further processing.

After the ZnO deposition and shipping, the substrates were moved into a nitrogen atmosphere without any further air exposure. Thermal evaporation of the subsequent layers was carried out in a vacuum chamber with a base pressure < $10^{-6}$ mbar. After the deposition of the organic layers ($\alpha$-NPD (Sigma Aldrich), BF-DPB (Lumtec) and BPAPF (Lumtec) all used as received) of different thicknesses the devices were finalized by $MoO_3$ (Lesker) and Al deposition through a shadow mask resulting in a pixel size of 0.16 $cm^2$. The thickness of the ZnO and the organic layers was measured with a Dektak[3ST] profilometer. Using a microscope cover glass from Menzel-Gläser the devices were sealed with a two-component epoxy glue from Huntsman Araldite 2011. The samples were stored in darkness and under nitrogen atmosphere. Intentional light exposure was performed by illumination with AM1.5G artificial sun light using a solar simulator (Newport 94042A). For the EL measurements, the bias was applied through a Keithley 2400 source-meter. The electric field within the organic layer was estimated by neglecting any voltage drop across the inorganic and assuming a built-in voltage of 1 V for all devices. This value results from the energy difference between the ZnO work function and the assumption of Fermi-level pinning slightly above the HOMO level of the organic, typically observed for Ohmic contacts.[34] All electroluminescence spectra were measured with an Andor SR393i-B spectrometer adjunct with a silicon detector DU420A-BR-DD and an InGaAs DU491A-1.7 detector. A calibrated Oriel 63355 lamp was used to determine the spectral response of the electroluminescence setup and to correct the acquired spectra. The conductivity of the ZnO films was measured on non-conductive glass substrates using interdigitated Au top-electrodes with a distance of 100 µm. The current flowing between the electrodes was measured under nitrogen atmosphere with a Keithley 2400 source-meter.

Ultraviolet photoelectron spectroscopy (UPS) experiments were performed at an Omicron Compact-based UHV vacuum system. Photoelectrons were excited using a HIS13 Helium gas discharge lamp $HeI_\alpha$ = 21.21 eV. Electrons were detected by an Omicron EA125 spectrometer with an energy resolution below 120 meV. The samples were biased to −10 V to measure work functions.

### III. Results

The EL spectra of BF-DPB deposited on SolGel-ZnO as well as $\alpha$−NPD and BPAPF deposited on ALD-ZnO show distinct, single peaks (see Figure 1) at similar driving conditions. By trend, their energies correlate with the shift of the ionization potentials and hence an expected increase of the hybrid CT emission energy. We note, that the exact energy landscape at the interface is unknown and that the reported literature values provide a rough estimate only.

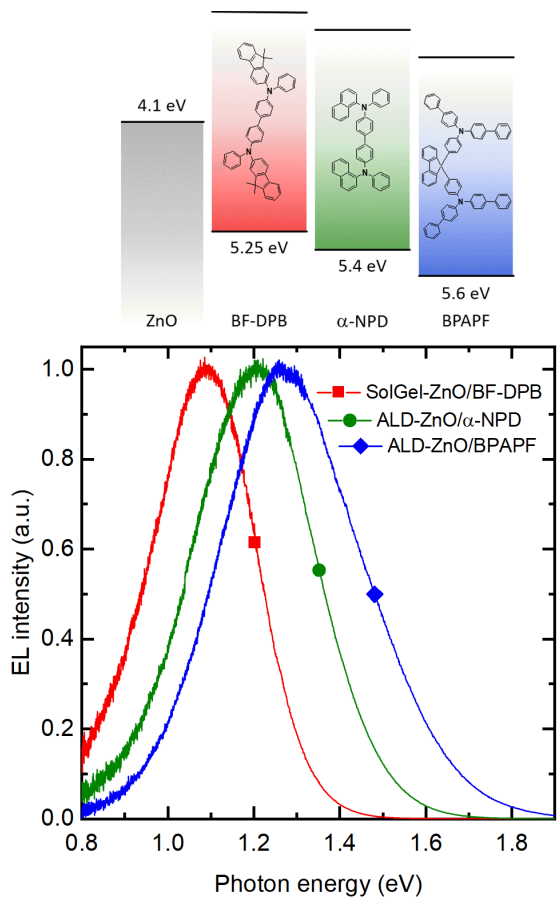

*Figure 1 Electroluminescence (EL) spectra of planar bilayer devices using BF-DPB, α-NPD and BPAPF as the donor. ZnO was deposited from solution (SolGel) and by Atomic Layer Deposition (ALD). The blue shift in the EL peak position correlates by trend with an increase of the interfacial energy gap. The cartoon shows the energy of the conduction band minimum of ZnO and the HOMO level of the organic small molecules.*

Moreover, the exact peak position is influenced by the applied bias during the EL measurement. This is illustrated in Figure 2 for a SolGel-ZnO/α-NPD device. While the EL signal is clearly dominated by the HCT feature, its spectral position blue shifts as the driving voltage is increased from 2.75 V to 6 V. Notably, the absolute peak position at a given voltage depends critically on the history of the sample. The spectra in Figure 2 a) were recorded before the sample was exposed to any UV light. Special care was taken to fabricate the device in a UV free environment. Figure 2 b) shows EL spectra of the same sample after 30 min of light soaking under AM1.5G conditions in a $N_2$ atmosphere. All spectra are clearly blue shifted compared to the pristine, UV protected sample. At the same time, the relative shift caused by the applied bias is reduced, as indicated by the vertical dashed lines. Finally, the sample was kept in darkness (without air exposure) for 18 h for relaxation from the light soaking. The EL spectra of the relaxed sample are shown in Figure 2 c). While all spectra exhibit a slight back shift towards lower energies, they are still far from the pristine state and the bias induced shift of the peak position remains virtually the same as immediately after UV exposure.

The effect of light soaking on the absolute and relative energy of the HCT signature becomes apparent from Figure 3 a), where the EL peak positions determined by Gaussian fits (grey, dashed lines in Figure 2) are plotted against the electric field. In the low field regime the peak energy scales linearly with the field. As indicated above, the dependency is stronger for the pristine device and the slope drops from about 2 eVnmV$^{-1}$ to approximately 1.2 eVnmV$^{-1}$ after the initial light soaking step and remains practically constant when the device is stored in darkness. It is important to note that the relaxation in darkness after the exposure to AM1.5G conditions is accompanied by a drastic change in the conductivity of the ZnO layer, as evident from Figure 4 a). This is attributed to a decay of photo-doping in ZnO.[35–37] The fact that the conductivity drop by two orders of magnitude for the relaxed device has no effect on the slope in Figure 3 a) justifies the assumption that the applied voltage drops solely across the organic layer, at least once the device has been exposed to UV light.

The situation is similar for the ALD-ZnO devices. This is illustrated in Figure 3 b) for an already light soaked ALD-ZnO hybrid device comprising α-NPD as the donor. Even though the conductivity change is less severe for the ALD-ZnO (red curve in Figure 4 a), it still drops by one order of magnitude within 24 h after the light soaking is stopped. We would like to point out that the sample in Figure 3 b) was not prepared under UV protected conditions and that the exact level of photo-doping is not known for our ALD-based devices. Nevertheless, the devices have been stored in darkness for at least 12 h, before the dark EL measurement. Judging from the absolute change in EL peak position upon 5 min additional light soaking under AM1.5G conditions the change in photo-doping appears to be significant, yet, the slope of the low field regime in Figure 3 b) remains basically unchanged.

The drastic difference in conductivity observed for the different levels of photo-doping would be expected to influence any present voltage drop across the metal oxide and thus to change the slope of the HCT energy as a function of the applied electric field. Against this prediction, the photo-doping of the ZnO leaves the field dependence virtually unchanged once an initial light soaking procedures is carried out and the mechanism proposed by Panda et al. [23] can clearly be ruled out for our light soaked devices.

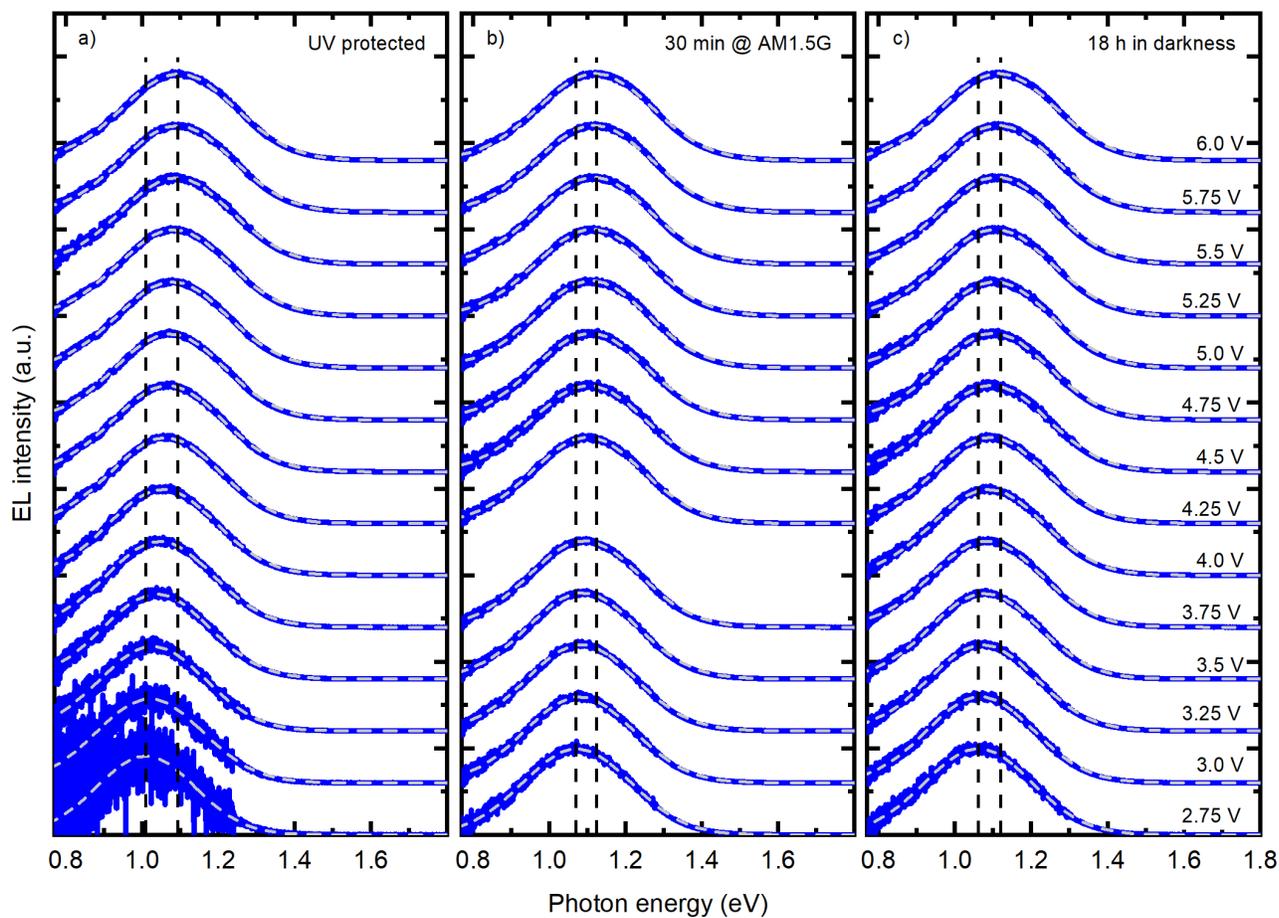

*Figure 2: Electroluminescence spectra of α-NPD (60 nm) deposited on SolGel-ZnO at different driving voltages from 2.75 V to 6 V. The same device was measured under three different conditions: a) Pristine device that has never seen UV-light even during production. b) Immediately after light soaking under AM1.5G conditions for 30 min. c) After storing the light soaked device in darkness for 18 h. By increasing the applied voltage, a blue shift of the HCT emission peak is observed. Each spectrum is fitted by a single Gaussian, as shown by the grey, dashed lines. The vertical dashed lines mark the photon energy of the HCT emission peak observed for the lowest and highest applied bias, respectively.*

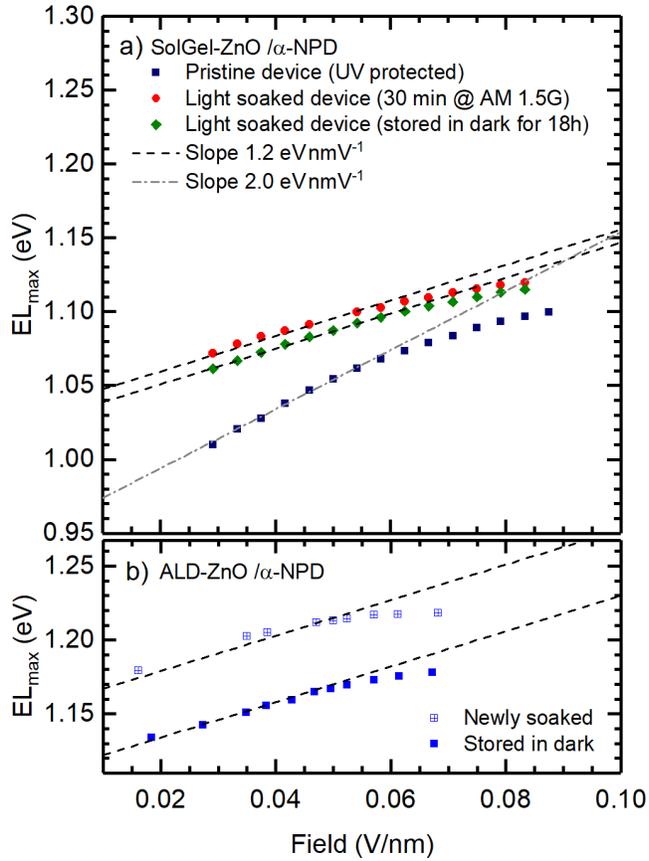

*Figure 3: a) Effect of initial light soaking on the field dependence of the EL peak position. A pronounced change of the slope is present for the very first exposure to UV light, afterwards it remains unaffected by the level of photo-doping. The conductivity change between the SolGel-ZnO of the freshly soaked (red) and relaxed (green) device is more than two orders of magnitude (see Figure 4). b) The phenomenon is similar for ALD-ZnO, where the slope of the freshly soaked device is maintained if the sample is left to relax in darkness over night. The electric field is estimated as the applied bias minus the built-in voltage, divided by the organic layer thickness.*

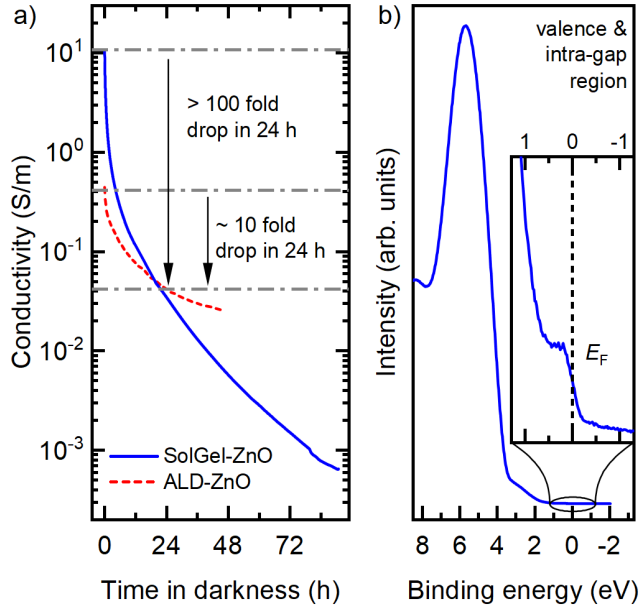

*Figure 4 a) Influence of the photo-doping decay on the conductivity of SolGel-ZnO and ALD-ZnO films. The samples have been exposed to AM1.5G conditions for 30 min before the light was turned off and the dark conductivity was traced. b) UPS valence band and intra-gap (inset) spectra of photo-doped SolGel-ZnO. Occupied states with the shape of the Fermi-Dirac distribution at the Fermi level $E_f$ evidence a metallic character of the surface.*

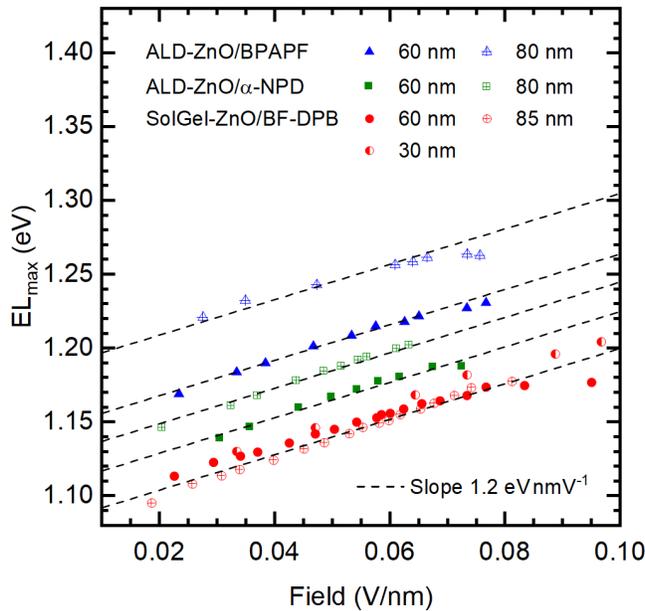

*Figure 5 Photon energy of the HCT emission peak for different organic layer thicknesses and molecules plotted against the internal electric field, estimated as the applied bias minus the built-in voltage, divided by the thickness of the organic layer. The dashed lines serve as a guide to the eye.*

In order to confirm that the observed EL peak shift is caused by the electric field rather than the absolute voltage, the thickness of the organic layer has been varied. The results are shown in Figure 5 for ALD-ZnO/α−NPD, ALD-ZnO/BPAPF and additionally SolGel-ZnO/BF-DPB hybrid devices. Clearly, the absolute value of the EL peak energy is subject to a relatively large sample to sample variation that is typical for ZnO based hybrid devices and possibly influenced by the background photo-doping caused by unintentional UV-light exposure.[38] Nevertheless, the relative peak position as a function of the electric field, i.e. the internal voltage normalized to the thickness of the organic layer, is the virtually identical for all employed molecules and independent of the organic layer thickness. This suggests that the cause of the observed linear shift is indeed the electric field.

## IV. Discussion

As mentioned above a field effect where a delocalized HCT state is confined by a triangular potential well was recently suggested to explain the observed EL peak shift.[21] The interfacial hybrid energy gap is increased by the ground state energy $E_0$ of a one dimensional, triangular quantum well. The field dependence of the latter can be approximated as

$$E_0 \approx \left(\frac{\hbar^2}{2m_{\text{eff}}}\right)^{1/3} \cdot \left(\frac{9}{8}\pi e\right)^{2/3} \cdot F^{2/3}, \qquad (1)$$

where $e$ is the elementary charge, $\hbar$ is the reduced Planck constant, $F$ is the electric field and $m_{\text{eff}}$ is the effective mass.[39] As illustrated by the solid green line in Figure 6 a) it is possible to describe our data with the quantum well model assuming an effective mass of 1.7 times the free electron mass $m_{\text{e}}$. Even though eq. 1 predicts a $F^{2/3}$ field dependence, the actual curve in Figure 6 a) is practically linear in the relevant field regime. The reason is that the wave function is already considerably localized by the relatively high effective mass (cf. Figure 6 b), impeding the additional confinement by the electric field.

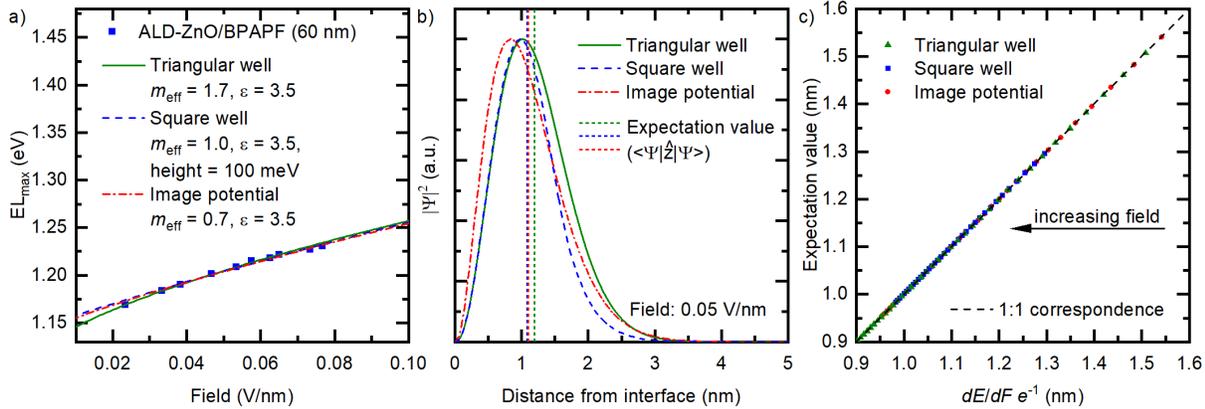

*Figure 6 Comparison of the three different considered quantum mechanical models. All models describe our data equally well (a) and the wave functions are similarly confined (b). The expectation value of the position operator corresponds directly to the derivative of the ground state energy with respect to the electric field for all three quantum mechanical models (c).*

Instead of a high effective mass we can also achieve the required degree of localization by considering a square well potential with the approximate width of one molecule (1.5 nm), infinite height at the interface, and an (arbitrary) height of 100 meV at the "border" of the molecule. This simple model also fits the data reasonably well, as shown by the dashed blue line in Figure 6 a). As for the triangular well, the expectation value amounts to slightly more than 1 nm at an intermediate field of 0.05V/nm (cf. Figure 6 b).

It is clear that a certain degree of localization is needed to model the experimental data. In the models with the triangular or rectangular wells, this localization is artificially tuned by the effective mass. In reality, the character of the ZnO surface has been shown to be strongly n-type or even metallic.[40–44] For our photo-doped SolGel-ZnO this is confirmed by the presence of occupied states (exhibiting the shape of the Fermi-Dirac-distribution) at the Fermi level in UPS measurements (see Figure 4 b). Hence, an alternative explanation for the hole localization is the Coulomb attraction of the hole by its image charge,[45]

$$V_{\text{image}} = - \frac{e^2}{16\,\pi\varepsilon\varepsilon_0 z} . \quad (2)$$

Here $\varepsilon$ is the relative permittivity of the organic layer, $\varepsilon_0$ is the vacuum permittivity and $z$ is the distance of the hole from the hybrid interface. In presence of an external field $F$ the Hamiltonian (with the momentum and position operators $\hat{p}$ and $\hat{z}$) is given by

$$\hat{H} = \frac{\hat{p}^2}{2m_{\text{eff}}} - \frac{e^2}{16\pi\varepsilon\varepsilon_0 \hat{z}} + eF\hat{z} . \quad (3)$$

This problem resembles the Stark effect of a hydrogen atom. However, since the image potential is screened by a factor of $\frac{1}{4\varepsilon}$, the contributions of the image charge and the external field to the electrostatic energy are of similar magnitude. The numerical solution provides a good agreement with the experimental data, yielding an effective hole mass of 0.7 $m_e$ (dash-dotted red line in Figure 6 a). The resulting wave function is again partially localized with an expectation value close to 1 nm (Figure 6 b).

It is obviously possible to describe the relative energy of our data with any of the three quantum mechanical models, if a suitable effective mass is chosen. In all cases, the bias affects the energy of the hybrid CT state by elevating the electrostatic energy of the hole. This additional energy is given by the interaction of the internal electric field with the dipole moment of the system, here given by the hole located at an effective distance from the hybrid interface. Consequently, regardless of the model, the local slope $\frac{dE}{dF}$ of the HCT emission energy $E$ in function of the applied electric field can be identified with the expectation value of the position operator which corresponds to the effective interface/hole distance in the scope of the discussed models. This is illustrated in Figure 6 c). We note that this identification yields distances which are on the order of the size of the long axis of the employed molecules. This appears overly large as it would imply that the holes were located on the far end of the molecule and that all molecules were highly ordered, standing upright at the hybrid interface. Qualitatively, however, the fact that for higher fields the interface/hole distance is reduced agrees with the observation that the slope decreases in the high field range. This implies that the charge is pushed towards the interface by the external field.

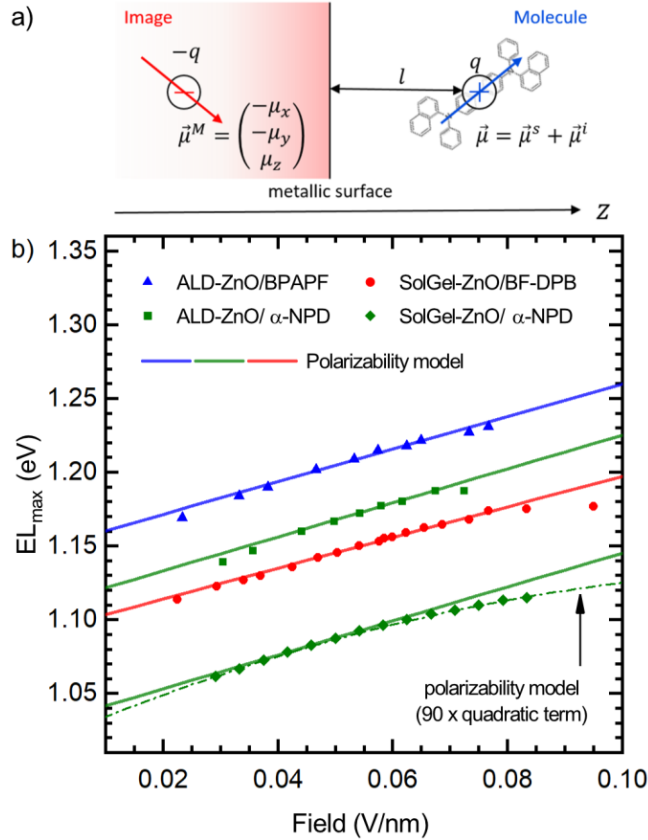

Figure 7 *Proposed model of a polarizable, molecular cation at a metallic surface interacting with its mirror image, illustrated in the cartoon in a) and applied to the measured data in b). The solid lines in b) are the result of the polarizability model for an equal number of flat lying and upright standing molecules (Δ=0.5), the dash-dotted line visualizes that the quadratic Stark effect can account for the saturation indicated at high fields. The absolute energy of all calculated curves has been adjusted by a suitable offset. The used values for $l_\parallel$ and $l_\perp$ are 1.0 nm and 0.5 nm for BPAPF, 1.0 nm and 0.2 nm for α-NPD, and 0.8 nm and 0.3 nm for BF-DPB, respectively.*

The predicted high localization implies that the quantum mechanical description can be approximated by a classical electrostatic expansion.[46] We therefore consider the molecular cation as a polarizable point multipole (see the cartoon in Figure 7 a) and, considering that the ZnO is heavily doped at the surface, approximate the ZnO film as a metal sheet. We can then treat the electrostatic interaction of the cation with its image in a perturbative way. The multipoles and polarizabilities we obtain from the optimized geometries of the cations in the gas phase, using the B3LYP functional and the 6-311g(d,p) basis set. We expand the electrostatic field around the molecule in multipoles and assume that all induction effects can be described by the polarizability tensor $\chi_{ij}$ of the positively charged molecule. The total electrostatic energy of the system can then be written as[46]

$$U = U_\text{charge-charge} + U_\text{charge-dipole} + U_\text{dipole-dipole} + U_\text{charge-quadrupole} + U_\text{induction} + U_\text{field}. \quad (4)$$

Assuming that the polarizability tensor is isotropic, only the $z$-component, i.e. the component parallel to the electric field $F$, of the induced dipole is relevant. Neglecting the dipole-dipole and charge-quadrupole interactions and assuming that the hole is located at the center of mass of the molecule and at distance $l$ from the interface, the total electrostatic energy of the system simplifies to

$$U = \underbrace{-\frac{1}{8\pi\epsilon_0 l}q^2}_{\text{charge-charge}} \underbrace{-\frac{2}{16\pi\epsilon_0 l^2}q\mu_z}_{\text{charge-dipole}} + \underbrace{\frac{1}{2}\chi^{-1}(\mu_z^i)^2}_{\text{induction}} \underbrace{-\mu_z F_z - qlF_z}_{\text{field}}, (5)$$

where $\mu_z = \mu_z^s + \mu_z^i$ and the superscripts $s$ and $i$ mark the static and induced dipoles, respectively and $2l$ is the distance between the cation and its mirror image. Here, $U_{\text{charge-charge}}$ is the interaction between the hole and its image charge, $U_{\text{charge-dipole}}$ accounts for the interaction between the hole and the image dipole $\mu_z^M$ as well as between the image charge and the molecular dipole $\mu_z$ (static as well as induced, cf. Figure 7 a). Note that the induction term $U_{\text{induction}}$ increases the interaction energy, penalizing the induced dipole, and is zero in the ground state. Finally, the field term $U_{\text{field}}$ accounts for the interaction of the external electric field with the total dipole moment of the system. Minimization of the total energy with respect to the induced dipole $\mu_z^i$ yields

$$\mu_z^i = \chi\left(F_z + \frac{2}{16\pi\epsilon_0 l^2}q\right). (6)$$

We note that the induced dipole has two contributions: one due to the external field and one due to the image charge that additionally polarizes the molecule. Substituting this back into the total energy we obtain

$$U = -\frac{1}{8\pi\epsilon_0 l}q^2 - \frac{2}{16\pi\epsilon_0 l^2}q\mu_z^s + \frac{1}{2}\chi\left(\frac{2}{16\pi\epsilon_0 l^2}q\right)^2 - \left(ql + \frac{2}{16\pi\epsilon_0 l^2}\chi q + \mu_z^s\right)F_z - \frac{1}{2}\chi F_z^2. (7)$$

The first three terms are field independent and will add a constant energy to the field-free hybrid CT energy. The linear in the electric field term describes the linear Stark effect and has three contributions: the interaction of the external field with (i) a positive point charge $q$ at distance $l$ from the interface, (ii) the dipole induced by the image potential, and (iii) the static dipole of the molecule. The quadratic in field term is the interaction of the external field-induced dipole with the external field itself. For the slope of the energy with the external field we then obtain

$$\frac{\partial U}{\partial F_z} = -ql - \frac{2}{16\pi\epsilon_0 l^2}\chi q - \mu_z^s - \chi F_z. (8)$$

This expression shows that the experimentally observed slope of the EL peak maximum vs electric field depends not only on the interface/hole distance: While the static dipole moment of all studied molecules is on the order of 1-2 Debye and can be ignored in our case, the molecular polarizability tensors of the cations (summarized in in Table 1) are considerable. Therefore, the first and second term are of similar magnitude and govern the field dependence in the low-field regime.

To account for the distribution of molecular orientations, we note that both the geometric shape and electrostatic response of the studied molecules are anisotropic. For axially symmetric molecules, the distance between the center of mass of the molecule and the metal oxide surface can be modeled as $l = l_\parallel + (l_\perp - l_\parallel)\sin\theta$ and the polarizability in field direction as $\chi = \chi_\perp \sin^2\theta + \chi_\parallel \cos^2\theta$. Here, $\theta$ is the angle between the long molecular axis and the surface normal. $l_\parallel$ and $l_\perp$ are the distances from the center of mass of the molecule to the oxide surface when the molecule is standing up ($\theta = 0°$) or lying flat ($\theta = $ 90°) on the surface. $\chi_\parallel$ and $\chi_\perp$ are the components of the polarization tensor parallel and perpendicular to the long molecular axis. Since the exact distribution of the molecular orientation at the hybrid junction is not known, we limit ourselves to a bimodal distribution of upright standing and flat lying molecules and estimate the average field dependence by assuming the fraction Δ of flat lying molecules. The expected slope of the EL peak energy in function of the electric field is then obtained by

$$\left\langle \frac{\partial U}{\partial F_z} \right\rangle_\theta \cong -q[l_\parallel + \Delta(l_\perp - l_\parallel)] - \frac{2}{16\pi\epsilon_0} q \left[ (1-\Delta)\frac{\chi_\parallel}{l_\parallel^2} + \Delta \frac{\chi_\perp}{l_\perp^2} \right] - \mu_z^S - [\chi_\parallel + \Delta(\chi_\perp - \chi_\parallel)]F_z . \quad (9)$$

Assuming an equal number of flat lying and standing molecules, i.e. $\Delta = 0.5$, and the values given in Table 1, we now use Equation (9) to describe the low-field regime in Figure 7 b). The Δ-weighted mean of the $l_\parallel$ and $l_\perp$ values yields average interface hole distances of about 0.6 nm for a-NPD and BF-DPB and about 0.8 nm for BPAPF.

While the linear regime of our data is accurately reproduced by Equation (9), the quadratic term, i.e. the interaction of the field-induced dipole with the field itself, is too weak to describe the observed downward bending at high fields. To improve the agreement, the dipole-dipole interaction term $U_{\text{dipole-dipole}} = -\frac{2}{32\pi\epsilon_0 l^3}\mu_z^2$ can additionally be taken into account. As a result, the prefactor of the quadratic in field term changes from $\frac{\chi}{2}$ to $\frac{1}{2}\frac{\chi}{1-2\chi T_{zz}}$, where $T_{zz} = \frac{2}{32\pi\epsilon_0 l^3}$. The enhancement of the quadratic term reduces the overall energy at high fields, which is observed experimentally in Figure 3. In order to describe our data in the entire field range, the necessary enhancement should be of the order of 100, as illustrated by the dashed-dotted green line in Figure 7 b). The interaction of the induced dipole with its image and with the field itself can in principle lead to such an enhancement. A quantitative estimate is, however, not possible: within our simple model the point-dipole approximation leads to a so-called induced-dipole divergence at small intermolecular separations. For strongly polarizable cations the onset of this divergence matches the distances found from the fitting of the low field regime, which correspond to average sizes of the employed molecules. Thus, the interaction of the induced dipole with its image can explain the energy saturation at high fields. More quantitative treatments with damped electrostatic interactions at short distances, such as the Thole model, are of course possible [47–49] but will not provide an additional physical insight into the problem.

Table 1 Calculated polarizabilities for the studied compound cations (in Bohr³). The xx component is along the long molecular axis.

| Compound | $\chi_{xx}$ | $\chi_{xy}$ | $\chi_{yy}$ | $\chi_{xz}$ | $\chi_{yz}$ | $\chi_{zz}$ |
|---|---|---|---|---|---|---|
| α-NPD | 2044 | 116 | 577 | -13 | 0 | 467 |
| BF-DPB | 3723 | 33 | 732 | -145 | -4 | 665 |
| BPAPF | 5146 | -142 | 1611 | 0 | 0 | 859 |

# V. Conclusion

We have presented a combined experimental and theoretical study towards understanding the origin of the bias dependence of the EL peak energy in hybrid metal oxide / organic semiconductor type II heterojunctions. We have demonstrated that initial exposure to simulated sunlight drastically changes the slope of the peak shift as a function of electric field, but that after this initial light soaking process, the slope becomes independent of the exact level of photo-doping despite the drastic change of the conductivity of ZnO. Combined with results for samples with different organic layer thickness and ZnO preparation conditions, we conclude that the blue-shift of the EL maximum with applied bias is related to the electric field inside the organic layer, and not caused by state filling. Instead, sensitive UPS measurements confirm the previously reported metallic character of ZnO surfaces for our photo-doped SolGel material. We show that different quantum mechanical models, including the previously proposed triangular quantum well, capture the observed trend if a suitable effective mass is chosen. However, for a disordered molecular semiconductor, as employed here, the effective mass lacks physical meaning and its sole purpose is the adjustment of the hole localization in the framework of an otherwise delocalized continuum model. We alternatively propose a classical electrostatic model which considers the interaction of a highly polarizable molecular cation with its mirror image and the electric field in the organic layer. From this we learn that the total dipole moment of the system is not simply given by the distance of the hole from the interface ($ql$), but has a significant contribution stemming from the dipole induced by the image charge ($\frac{2}{16\pi\epsilon_0 l^2}\chi q$). The latter in turn is the result of the large polarizabilities of molecular cations that are approximately two orders of magnitude larger than the polarizabilities of the corresponding neutral molecules. Since the static dipole moment is negligible for the employed molecules, these two contributions are responsible for the slope of the linear Stark effect in the low field regime. We also propose that the quadratic Stark effect causes the energy shift to saturate at high fields, and that the interaction of the induced dipole with its image dipole contributes significantly to this effect. However, again because of the large polarizabilities, the point dipole approximation breaks down at larger fields and does not allow a fully quantitative analysis.

Finally, we would like to point out that the slope of the EL peak shift, reported for sputtered ZnO/CBP samples by Panda et al. and attributed to state filling, is significantly lower than what we have observed here. Together with our observation that the EL-peak shift behavior changes upon initial light soaking, this indicates that the mechanism causing the shift critically depends on the properties of the metal oxide, its preparation conditions and sample history.

# Acknowledgement

We thank Koen Vandewal (Hasselt University) for providing materials and helpful advice. This project was supported by the German Research Foundation (DFG) within the collaborative research center 951 "Hybrid Inorganic/Organic Systems for Opto-Electronics (HIOS)". U.H. and F.P. acknowledge the DFG for funding. D.A. thanks the BMBF grant InterPhase (FKZ 13N13661) and the European Union Horizon 2020 research and innovation program 'Widening materials models', Grant Agreement No. 646259 (MOSTOPHOS).

## Author contribution

U.H. and S.Z. contributed equally to this work.